\documentclass[aps,pra,preprintnumbers,showpacs,tightenlines]{revtex4}
\usepackage{amssymb}
\usepackage{amsmath}
\usepackage{graphicx}
\usepackage{epsfig}
\usepackage{subfigure}
\usepackage{amsfonts}
\usepackage{CJK}

\begin{document}

\title{Quantum logical gates with four-level SQUIDs
coupled to a superconducting resonator}

\author{Xiao-Ling He$^{1}$, Chui-Ping Yang$^{2}$, Sheng Li$^{3}$, Jun-Yan Luo$^{1}$, and Siyuan Han$^{4}$}

\address{$^1$School of Science, Zhejiang University of Science and
Technology, Hangzhou, Zhejiang 310023, China}

\address{$^2$Department of Physics, Hangzhou Normal University,
Hangzhou, Zhejiang 310036, China}

\address{$^3$Department of Physics, Zhejiang Normal University,
Jinhua, Zhejiang 321004, China}

\address{$^4$Department of Physics and Astronomy, University of
Kansas, Lawrence, Kansas 66045, USA}

\date{\today}

\begin{abstract}
We propose a way for realizing a two-qubit controlled phase gate
with superconducting quantum interference devices (SQUIDs) coupled
to a superconducting resonator. In this proposal, the two lowest
levels of each SQUID serve as the logical states and two
intermediate levels of each SQUID are used for the gate
realization. We show that neither adjustment of SQUID level
spacings during the gate operation nor uniformity in SQUID
parameters is required by this proposal. In addition, this
proposal does not require the adiabatic passage or a second-order
detuning and thus the gate is much faster.
\end{abstract}

\pacs{03.67.Lx, 42.50.Dv, 85.25.Dq} \maketitle

\textit{Introduction.}---Superconducting devices including cooper
pair boxes, Josephson junctions, and superconducting quantum
interference devices (SQUIDs) have appeared to be among the most
promising candidates for scalable quantum computing, due to design
flexibility, large scale integration, and compatibility to
conventional electronics [1-3]. In the past few years, for SQUID
systems, many theoretical methods for realizing a single-qubit
gate and a two-qubit controlled-phase (CP) or controlled-NOT
(CNOT) gate have been presented [4-15].

For realizing a two-qubit CP gate with SQUIDs, several methods
have been proposed based on cavity QED technique [8-15]. These
methods are of importance in building quantum logic gates and open
a new avenue for the physical realization of quantum information
processing with SQUIDs in cavity QED. However, we note that these
methods have some disadvantages. For instances: (i) the methods
presented in [8,9] require adjustment of the level spacings of
SQUIDs during the gate operation, thus decoherence caused due to
the adjustment of level spacings may pose a severe problem; (ii)
the methods proposed in [10,11] require slowly changing the Rabi
frequencies to satisfy the adiabatic passage and the approaches
introduced in [12-14] require a second-order detuning to achieve
an off-resonant Raman coupling between two relevant levels; note
that when the adiabatic passage or a second-order detuning is
applied, the gate becomes slow (the gate time is on the order of
one microsecond to a few microseconds [11,13]); and (iii) the
proposal reported in [15] employs a two-mode resonator/cavity as
well as a second-order detuning between the two cavity modes;
technically speaking, the requirement for a SQUID interacting with
more than one cavity or resonator modes is difficult to meet. In
addition, it is noted that although two-qubit CNOT, CP, or iSWAP
gates have been experimentally demonstrated in superconducting
charge qubits, flux qubits, and phase qubits [16-18], to the best
of our knowledge, no experimental demonstration of a two-qubit
gate with SQUID qubits in cavity QED has been reported.

In this paper, we present an alternative method for implementing a
two-qubit CP gate with two SQUIDs coupled to a superconducting
resonator. As shown below, this proposal has the following
advantages: (a) there is no need for adjusting the level spacings
of SQUIDs during the gate operation, thus decoherence caused by
tuning the SQUID level spacings is avoided; (b) neither slowly
changing the Rabi frequency nor the use of second-order detuning
is required, thus the gate is significantly faster (as shown
below, the operation time of the gate is on the order of ten
nanoseconds); and (d) only one mode of the resonator is employed.
In addition, this proposal does not require identical coupling
constants of each SQUID with the resonator and thus is tolerable
to inevitable nonuniformity in device parameters. We believe that
this work is of interest because it avoids most of the problems
existing in the previous proposals.

\textit{Basic theory.}---The SQUIDs considered throughout this paper are rf
SQUIDs each consisting of a Josephson tunnel junction enclosed by a
superconducting loop. 
The Hamiltonian for an rf SQUID, with junction capacitance $C$ and loop
inductance $L$, can be written in the usual form [19]
\begin{equation}
H_s=\frac{Q^2}{2C}+\frac{\left( \Phi -\Phi _x\right) ^2}{2L}-E_J\cos \left(
2\pi \frac \Phi {\Phi _0}\right) ,
\end{equation}
where $\Phi $, the magnetic flux threading the ring, and $Q$, the total
charge on the capacitor, are the conjugate variables of the system, %
$\Phi _x$ is the static (or quasistatic) external magnetic flux applied to
the ring, and $E_J$ $\equiv I_c\Phi _0/2\pi $ is the Josephson coupling
energy, where $I_c$ is the critical current of the junction and $\Phi
_0=h/2e $ is the flux quantum.

\begin{figure}[tbp]
\includegraphics[bb=118 362 484 577, width=8.6 cm, clip]{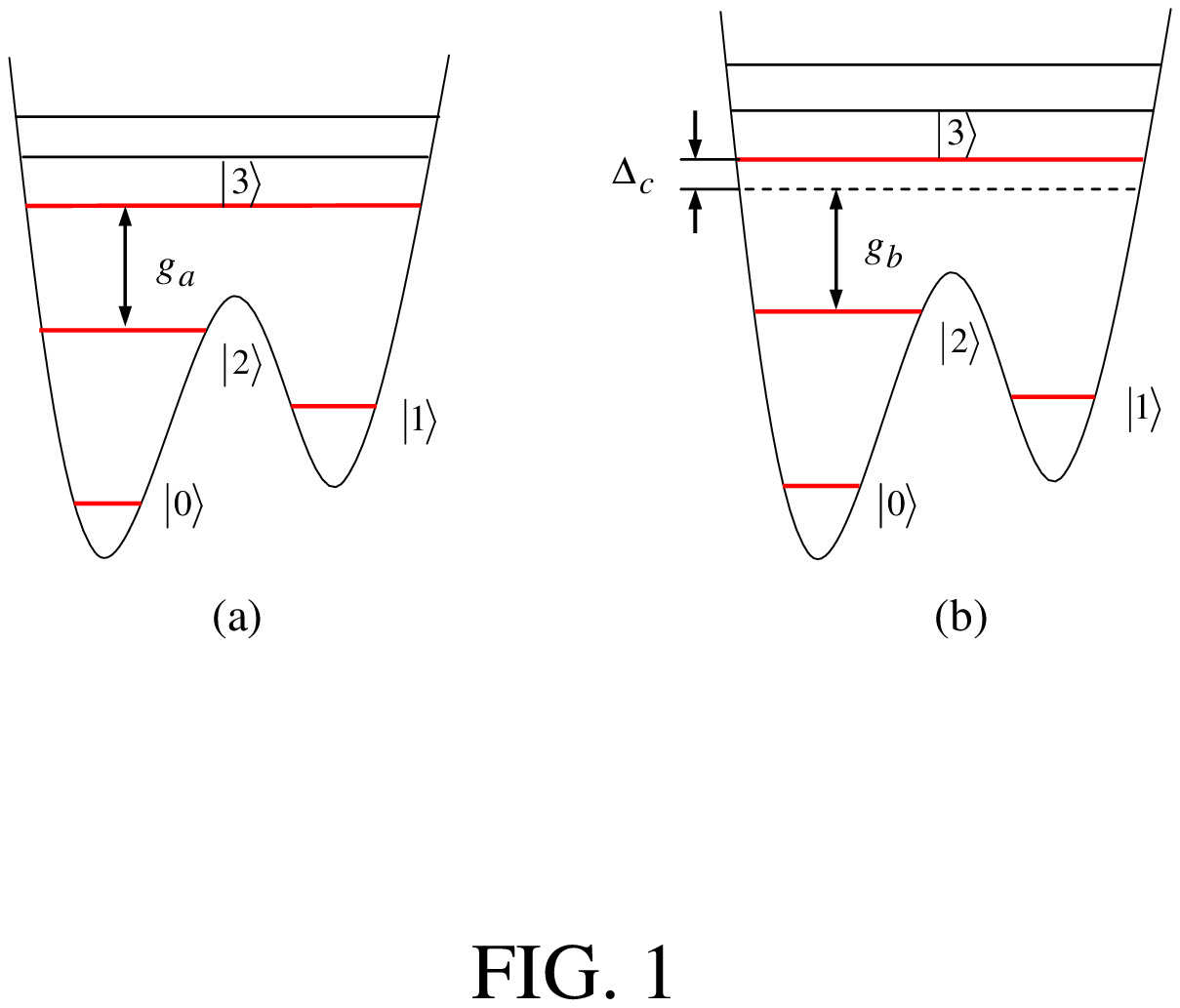} %
\vspace*{-0.08in}
\caption{(a) SQUID-resonator resonant interaction. (b) SQUID-resonator
off-resonant interaction. The difference between the level spacings in (a)
and the level spacings in (b) can be achieved by choosing different device
parameters for SQUIDs.}
\label{fig:1}
\end{figure}

\textit{A). SQUID-resonator resonant interaction.} Consider a SQUID (say
SQUID $a$) coupled to a single-mode resonator and driven by a classical
microwave pulse. The SQUID is biased properly to have four lowest levels,
which are denoted by $\left| 0\right\rangle ,$ $\left| 1\right\rangle ,$ $%
\left| 2\right\rangle ,$ and $\left| 3\right\rangle ,$ respectively
[Fig.~1(a)]. The resonator mode is resonant with the $\left| 2\right\rangle
\leftrightarrow \left| 3\right\rangle $ transition but decoupled (highly
detuned) from the transition between any other two levels, which can be
readily achieved by adjusting level spacings of the SQUID [8,20]. In the
interaction picture, the interaction Hamiltonian for the SQUID and the
resonator mode, after making the rotating-wave approximation, can be written
as [8]
\begin{equation}
H_I=\hbar \left( g_ac^{+}\left| 2\right\rangle _a\left\langle 3\right| +%
\text{H.c.}\right) .
\end{equation}
Here, the subscript $a$ represents SQUID $a;$ $c^{+}$ and $c$ are the photon
creation and annihilation operators of the resonator mode with frequency $%
\omega _c$; $g_a$ is the coupling constant between the resonator mode and
the $\left| 2\right\rangle \leftrightarrow \left| 3\right\rangle $
transition of SQUID $a$. The initial state $\left| 3\right\rangle _a\left|
0\right\rangle _c$ and $\left| 2\right\rangle _a\left| 1\right\rangle _c$ of
the system, under the Hamiltonian (2), evolve as follows
\begin{eqnarray}
\left| 3\right\rangle _a\left| 0\right\rangle _c &\rightarrow &\cos
g_at\left| 3\right\rangle _a\left| 0\right\rangle _c-i\sin g_at\left|
2\right\rangle _a\left| 1\right\rangle _c,  \nonumber \\
\left| 2\right\rangle _a\left| 1\right\rangle _c &\rightarrow &\cos
g_at\left| 2\right\rangle _a\left| 1\right\rangle _c-i\sin g_at\left|
3\right\rangle _a\left| 0\right\rangle _c,
\end{eqnarray}
where $\left| 0\right\rangle _c$ and $\left| 1\right\rangle _c$ are the
vacuum state and the single-photon state of the resonator mode, respectively.

\textit{B). SQUID-resonator off resonant interaction.} Consider a
system composed of a SQUID (say SQUID $b$) and a single-mode
resonator. Suppose that the resonator mode is off-resonant with
the $\left| 2\right\rangle \leftrightarrow \left| 3\right\rangle $
transition (i.e., $\Delta _c=\omega _{32}-\omega _c\gg g_b$) while
decoupled from the transition between any other two levels of
SQUID $b$ [Fig. 1(b)]. Here, $\Delta _c$ is the detuning between
the $\left|
2\right\rangle \leftrightarrow \left| 3\right\rangle $ transition frequency $%
\omega _{32}$ of SQUID $b$ and the resonator mode frequency
$\omega _c,$ and $g_b$ is the coupling constant between the
resonator mode and the $\left| 2\right\rangle \leftrightarrow
\left| 3\right\rangle $ transition. The effective interaction
Hamiltonian in the interaction picture can be written as [21,22]
\begin{equation}
H_e=\hbar \frac{g_b^2}{\Delta _c}\left( \left| 3\right\rangle _b\left\langle
3\right| -\left| 2\right\rangle _b\left\langle 2\right| \right) c^{+}c,
\end{equation}
where the subscript $b$ represents SQUID $b.$

From the Hamiltonian (4), it is straightforward to see that if the resonator
mode is initially in a single-photon state $\left| 1\right\rangle _c$, the
time evolution of the states of the system is then given by
\begin{eqnarray}
\left| 2\right\rangle _b\left| 1\right\rangle _c &\rightarrow
&e^{ig_b^2t/\Delta _c}\left| 2\right\rangle _b\left| 1\right\rangle _c,
\nonumber \\
\left| 3\right\rangle _b\left| 1\right\rangle _c &\rightarrow
&e^{-ig_b^2t/\Delta _c}\left| 3\right\rangle _b\left| 1\right\rangle _c,
\end{eqnarray}
which introduces a phase shift $e^{ig_b^2t/\Delta _c}$ to the state $\left|
2\right\rangle $ while $e^{-ig_b^2t/\Delta _c}$ to the state $\left|
3\right\rangle $ of the SQUID, when the resonator mode is in the state $%
\left| 1\right\rangle _c.$ Note that the states $\left| 2\right\rangle
_b\left| 0\right\rangle _c$ and $\left| 3\right\rangle _b\left|
0\right\rangle _c$ remain unchanged under the Hamiltonian (4).

In the following gate operations, we will need this resonant
interaction between the pulse and SQUIDs. Note that the resonant
interaction between the pulse and the SQUIDs can be completed
within a very short time, by increasing the pulse Rabi frequency
(i.e., by increasing the intensity/amplitude of the pulse).

\textit{Two-qubit CP gate.}---Let us consider two SQUIDs $a$ and
$b.$ By choosing different device parameters for each SQUID,
SQUIDs $a$ and $b$ can have the four-level configurations as
depicted in Fig. 1(a) and Fig. 1(b), respectively. The two logic
states of a SQUID qubit are represented by the two lowest levels
$\left| 0\right\rangle $ and $\left| 1\right\rangle ,$ while the
two intermediate levels $\left| 2\right\rangle $ and $\left|
3\right\rangle $ of each SQUID are utilized for the gate
realization. For the notation convenience, we here denote the
ground state (the first excited state) as level $\left|
1\right\rangle $ ($\left|
0\right\rangle $) for SQUID $b$ [Fig. 2(a$^{\prime }$,b$^{\prime }$,c$%
^{\prime }$,d$^{\prime }$,e$^{\prime }$)]. We suppose that the resonator
mode is resonant with $\left| 2\right\rangle \leftrightarrow \left|
3\right\rangle $ transition of SQUID $a$ while off-resonant with the $\left|
2\right\rangle \leftrightarrow \left| 3\right\rangle $ transition of SQUID $b
$, which can be reached by prior adjustment of the level spacings of SQUIDs $%
a$ and $b$. In addition, we assume that the resonator mode is initially in
the vacuum state $\left| 0\right\rangle _c.$ The notations $\omega
_{31}^{(i)},$ $\omega _{20}^{(i)},$ and $\omega _{21}^{(i)}$ involved in the
following gate operations are the $\left| 1\right\rangle \leftrightarrow
\left| 3\right\rangle $ transition frequency, the $\left| 0\right\rangle
\leftrightarrow \left| 2\right\rangle $ transition frequency, and the $%
\left| 1\right\rangle \leftrightarrow \left| 2\right\rangle $ transition
frequency of SQUID $i$ ($i=a,b$).

\begin{figure}[tbp]
\includegraphics[bb=103 346 447 744, width=8.6 cm, clip]{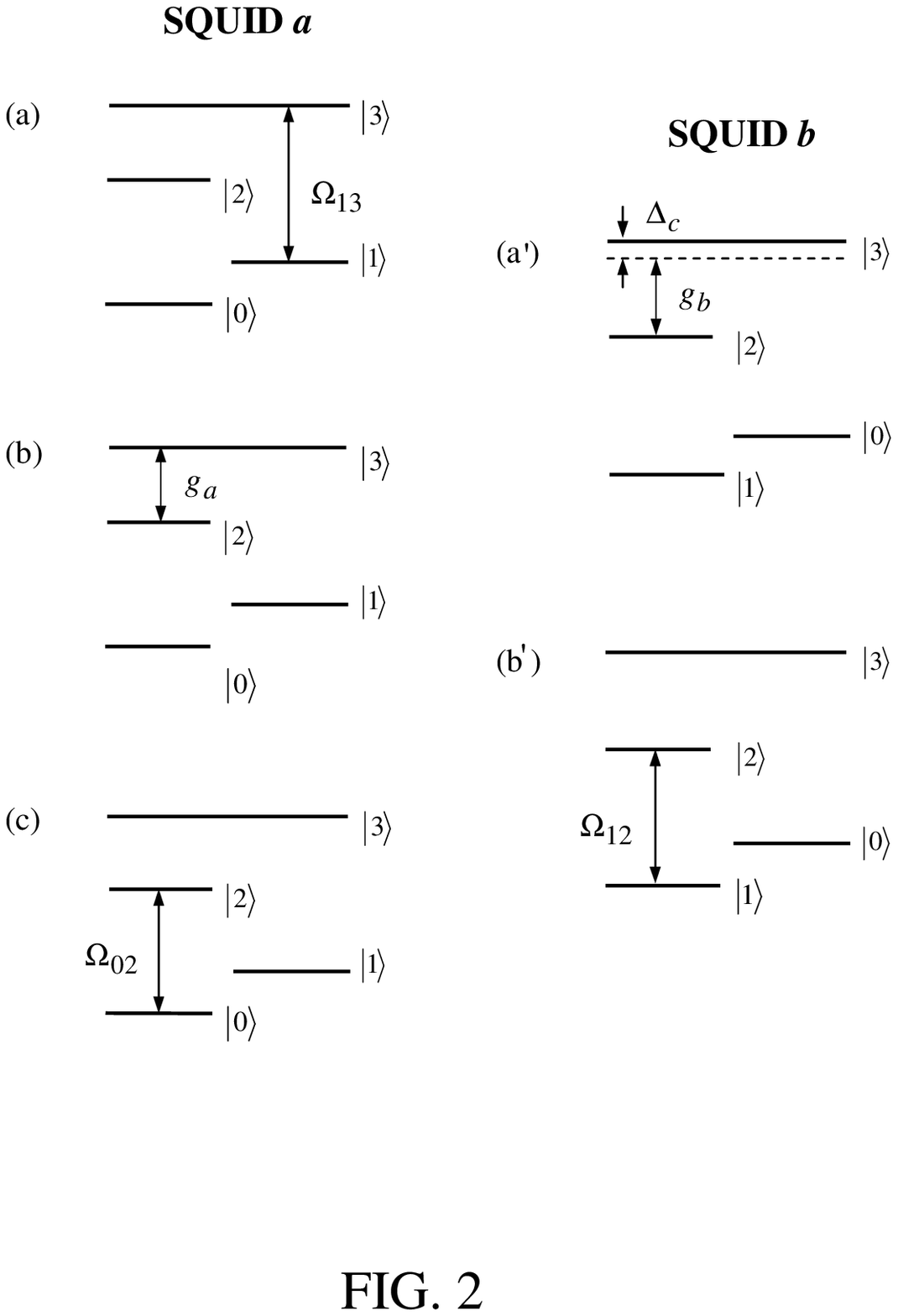} %
\vspace*{-0.08in}
\caption{Illustration of SQUIDs interacting with the resonator mode and/or
the microwave pulses during the gate performance. The figures on the left
(right) side correspond to SQUID $a$ ($b$).}
\label{fig:2}
\end{figure}

The operations for realizing a two-qubit CP gate are listed as
follows:

Step (i): Apply a microwave pulse (with a frequency $\omega _{\mu w}=\omega
_{31}^{(a)}$ and a phase $\phi =\pi $) to SQUID $a$ for a time interval $%
t_1=\pi /\left( 2\Omega _{13}\right) $ [Fig. 2(a)], to transform the state $%
\left| 1\right\rangle _a$ to $i\left| 3\right\rangle _a.$ Then,
wait for a time interval $t_1^{\prime }=\pi /\left( 2g_a\right) $
during which the $\left| 2\right\rangle \leftrightarrow \left|
3\right\rangle $ transition of SQUID $a$ resonantly interacts with
the resonator mode [Fig. 2(b)], to transform the state $\left|
3\right\rangle _a\left| 0\right\rangle _c$ to $-i\left|
2\right\rangle _a\left| 1\right\rangle _c,$ as shown in Eq. (3).

It can be found that after this step, the following transformation is
obtained:

\begin{equation}
\left| 1\right\rangle _a\left| 0\right\rangle _c\stackrel{\text{after }t_1}{%
\rightarrow }i\left| 3\right\rangle _a\left| 0\right\rangle _c\stackrel{%
\text{ after }t_1^{\prime }}{\rightarrow }\left| 2\right\rangle _a\left|
1\right\rangle _c.
\end{equation}
On the other hand, the state $\left| 0\right\rangle _a\left| 0\right\rangle
_c$ remains unchanged.

Step (ii): Apply a microwave pulse (with a frequency $\omega _{\mu
w}=\omega _{20}^{(a)}$ and a phase $\phi =\pi /2$) to SQUID $a$
[Fig. 2(c)] while a microwave pulse (with a frequency $\omega
_{\mu w}=\omega _{21}^{(b)}$ and a phase $\phi =-\pi /2$) to SQUID
$b$ [Fig. 2(b$^{\prime }$)]. The Rabi frequency for the pulse
applied to SQUID $a$ is $\Omega _{02}$ while the
Rabi frequency of the pulse applied to SQUID $b$ is $\Omega _{12}$. We set $%
\Omega _{02}=\Omega _{12},$ which can be achieved by adjusting the
intensities of the two pulses. After the pulse duration $t_2=\pi /(2\Omega
_{02})=\pi /(2\Omega _{12}),$ the state $\left| 2\right\rangle $ ($\left|
0\right\rangle $) of SQUID $a$ is transformed to the state $\left|
0\right\rangle $ ($-\left| 2\right\rangle $) while the state $\left|
1\right\rangle $ of SQUID $b$ is transformed to the state $\left|
2\right\rangle $.

Step (iii): Wait for a time $t_{3.}$ Note that in the case when
the resonator mode is in the photon state $\left| 1\right\rangle
_c,$ the levels $\left| 2\right\rangle $ and $\left|
3\right\rangle $ of SQUID $a$ are not populated after the above
operations. Therefore, there is no coupling between the resonator
mode and SQUID $a$. The resonator mode is off-resonant with the
$\left| 2\right\rangle \leftrightarrow \left| 3\right\rangle $
transition of SQUID $b$ [Fig. 2(a$^{\prime }$)]. It can be seen
from Eq. (5) that for $t_3=\pi \Delta _c/g_b^2,$ the state $\left|
2\right\rangle _b\left| 1\right\rangle _c$ changes to $-\left|
2\right\rangle _b\left| 1\right\rangle _c$. On the other hand, the
state $\left| 0\right\rangle _b\left| 0\right\rangle _c,\left|
0\right\rangle _b\left| 1\right\rangle _c,$ and $\left|
2\right\rangle _b\left| 0\right\rangle _c$ remain unchanged.

Step (iv): Apply a microwave pulse (with a frequency $\omega _{\mu
w}=\omega _{20}^{(a)}$ and a phase $\phi =-\pi /2$) to SQUID $a$
[Fig.~2(c)] while a microwave pulse (with a frequency $\omega
_{\mu w}=\omega _{21}^{(b)}$ and a phase $\phi =\pi /2$) to SQUID
$b$ [Fig.~2(b$^{\prime }$)]. Like step (ii), we set $\Omega
_{02}=\Omega _{12}$. After the pulse duration $t_2$ given in step
(ii), the state $\left| 0\right\rangle $ ($\left| 2\right\rangle
$) of SQUID $a$ is transformed to the state $\left| 2\right\rangle
$ ($-\left| 0\right\rangle $) while the state $\left|
2\right\rangle $ of SQUID $b$ is transformed back to the state
$\left| 1\right\rangle $.

Step (v): Perform an inverse operation of step (i) [Fig. 2(a,b)].
That is, wait for a time interval $t_1^{\prime }$ given in step
(i), during which the $\left| 2\right\rangle \leftrightarrow
\left| 3\right\rangle $ transition of SQUID $a$ resonantly
interacts with the resonator mode; and then apply a microwave
pulse (with a frequency $\omega _{\mu w}=\omega _{31}^{(a)}$ and a phase $%
\phi =\pi $) to SQUID $a$ for a time interval $t_1$ given in step (i). It
can be verified that after this step, the following transformation is
achieved:
\begin{equation}
\left| 2\right\rangle _a\left| 1\right\rangle _c\stackrel{\text{after }%
t_1^{\prime }}{\rightarrow }-i\left| 3\right\rangle _a\left| 0\right\rangle
_c\stackrel{\text{after }t_1}{\rightarrow }\left| 1\right\rangle _a\left|
0\right\rangle _c.
\end{equation}
On the other hand, the state $\left| 0\right\rangle _a\left| 0\right\rangle
_c$ remains unchanged.

The states of the whole system after each step of the above operations are
summarized in the following table:
\begin{eqnarray}
&&
\begin{array}{c}
\left| 00\right\rangle \left| 0\right\rangle _c \\
\left| 01\right\rangle \left| 0\right\rangle _c \\
\left| 10\right\rangle \left| 0\right\rangle _c \\
\left| 11\right\rangle \left| 0\right\rangle _c
\end{array}
\stackrel{\text{Step(i)}}{\longrightarrow }
\begin{array}{c}
\left| 00\right\rangle \left| 0\right\rangle _c \\
\left| 01\right\rangle \left| 0\right\rangle _c \\
\left| 20\right\rangle \left| 1\right\rangle _c \\
\left| 21\right\rangle \left| 1\right\rangle _c
\end{array}
\stackrel{\text{Step(ii)}}{\longrightarrow }
\begin{array}{c}
\left| 20\right\rangle \left| 0\right\rangle _c \\
\left| 22\right\rangle \left| 0\right\rangle _c \\
\left| 00\right\rangle \left| 1\right\rangle _c \\
\left| 02\right\rangle \left| 1\right\rangle _c
\end{array}
\stackrel{\text{Step(iii)}}{\longrightarrow }
\begin{array}{c}
\left| 20\right\rangle \left| 0\right\rangle _c \\
\left| 22\right\rangle \left| 0\right\rangle _c \\
\left| 00\right\rangle \left| 1\right\rangle _c \\
-\left| 02\right\rangle \left| 1\right\rangle _c
\end{array}
\nonumber \\
&&\stackrel{\text{Step(iv)}}{\longrightarrow }
\begin{array}{c}
\left| 00\right\rangle \left| 0\right\rangle _c \\
\left| 01\right\rangle \left| 0\right\rangle _c \\
\left| 20\right\rangle \left| 1\right\rangle _c \\
-\left| 21\right\rangle \left| 1\right\rangle _c
\end{array}
\stackrel{\text{Step(v)}}{\longrightarrow }
\begin{array}{c}
\left| 00\right\rangle \left| 0\right\rangle _c \\
\left| 01\right\rangle \left| 0\right\rangle _c \\
\left| 10\right\rangle \left| 0\right\rangle _c \\
-\left| 11\right\rangle \left| 0\right\rangle _c
\end{array}
,
\end{eqnarray}
where $\left| kl\right\rangle $ is abbreviation of the state
$\left| k\right\rangle _a\left| l\right\rangle _b$ of SQUIDs
($a,b$) with $k,l\in \{0,1,2\}$. It can be concluded from Eq.~(8)
that a two-qubit CP gate was achieved with two SQUIDs (i.e., the
control SQUID $a$ and the target SQUID $b$) after the above
process.

From the description above, it can be found that: (i) In contrast to the
previous proposals [8,9], the method presented above does not require
adjustment of the level spacings of the SQUIDs during the gate operation;
(ii) This method does not require slow variation of the Rabi frequency in
contrast to [10,11]; (iv) Compared with the previous approaches [12-14],
this method does not require a finite second-order detuning $\delta =\Delta
_c-\Delta _{\mu w}$ and thus the gate speed is improved by one order (here $%
\Delta _{\mu w}$ is the detuning of the pulse frequency with the transition
frequency between the two associated levels of SQUIDs; for the details, see
[12-14]); and (v) this method employs only one mode of the resonator, which
is different from the previous proposal [15].

\textit{Discussion.}---Let us give a brief estimate on the gate time. As
shown above, the total operation time is
\begin{equation}
\tau =2t_1+2t_1^{\prime }+2t_2+t_3=\pi /g_a+\pi \Delta _c/g_b^2+\pi /\Omega
_{13}+\pi /\Omega _{02},
\end{equation}
where $\Omega _{02}$ is equal to $\Omega _{12}$ (see steps (ii) and (iv)
above). Without loss of generality, let us consider $g_a\sim g_b$ $\sim
3.0\times 10^9$ s$^{-1}$, which is available at present [9]. By choosing $%
\Delta _c=10g_b,$ $\Omega _{13}\sim \Omega _{02}\sim 10g_a,$ we have $\tau
\sim 12$ ns.

Several issues related to the gate operations above need to be addressed as
follows:

\textit{i) }The level $\left| 3\right\rangle $ of SQUID $a$ is occupied in
steps (i) and (v). Since only SQUID-pulse resonant interaction and
SQUID-resonator resonant interaction are used in steps (i) and (v), the
operation time $t_1+t_1^{\prime }$ in step (i) or (v), equal to $\pi /\left(
2\Omega _{13}\right) +\pi /\left( 2g_a\right) ,$ can be significantly
shorten by increasing the pulse Rabi frequency $\Omega _{13}$ and the
coupling constant $g_a.$ Alternatively, one can design the SQUID $a$ to have
a sufficiently long energy relaxation time $\gamma _3^{-1}$ for the level $%
\left| 3\right\rangle .$ By doing these, we can have $\gamma _3^{-1}\gg
t_1+t_1^{\prime },$ such that decoherence caused by the energy relaxation of
the level $\left| 3\right\rangle $ of SQUID $a$ is negligibly small.

\textit{ii)} The occupation probability $p_3$ of the level $\left|
3\right\rangle $ for SQUID $b$ during step (iii) is given by [12]
\begin{equation}
p_3\simeq \frac{4g_b^2}{4g_b^2+\Delta _c^2},
\end{equation}
which need to be negligibly small in order to reduce the gate error. For the
choice of $\Delta _c=10g_b$, we have $p_3\sim 0.04,$ which can be further
reduced by increasing the ratio of $\Delta _c/g_b.$

\textit{iii}) For steps (i), (ii), (iv) and (v), the resonant interaction
between the resonator mode and the $\left| 2\right\rangle \leftrightarrow
\left| 3\right\rangle $ transition of SQUID $a,$ involved during the
application of the pulse, is unwanted. To minimize the effect of this
unwanted interaction on the gate, the Rabi frequencies $\Omega _{13}$ and $%
\Omega _{02}$ require to be much larger than the coupling constant $g_a,$
i.e., $\Omega _{13},\Omega _{02}\gg g_a.$ Note that this condition can be
achieved by increasing $\Omega _{13}$ and $\Omega _{02}$ (i.e., via
increasing the pulse intensity).

\textit{iv}) For either step (ii) or step (iv), when the SQUID $b$ is in the
state $\left| 2\right\rangle $ and the resonator mode is in the
single-photon state $\left| 1\right\rangle _c,$ the unwanted off-resonant
interaction between the resonator mode and the $\left| 2\right\rangle
\leftrightarrow \left| 3\right\rangle $ transition of SQUID $b$ induces a
phase shift $e^{ig_b^2t_2/\Delta _c}$ to the state $\left| 2\right\rangle $
of SQUID $b$, which will affect the desired gate performance. The effect of
this unwanted SQUID-resonator off-resonant interaction on the gate can be
made negligibly small as long as the condition $\Omega _{12}\gg g_b^2/\Delta
_c$ is met. In the following, we will give a discussion on the effect of
this unwanted interaction on the fidelity of the gate.

Suppose that the two SQUID qubits are initially in a generic state described
by $\left| \psi \left( 0\right) \right\rangle =\alpha \left| 00\right\rangle
+\beta \left| 01\right\rangle +\gamma \left| 10\right\rangle +\theta \left|
11\right\rangle $, where the coefficients satisfy the normalization. In the
ideal case, it can be seen from Eq. (10) that after the five-step operations
described above, the state $\left| \psi \left( 0\right) \right\rangle $
becomes $\left| \psi _{id}\left( \tau \right) \right\rangle =\alpha \left|
00\right\rangle +\beta \left| 01\right\rangle +\gamma \left| 10\right\rangle
-\theta \left| 11\right\rangle .$ On the other hand, when the effect of the
off-resonant interaction between the resonator mode and SQUID $b$ is
included during steps (ii) and (iv), one can easily work out the expression
for the final state $\left| \psi \left( \tau \right) \right\rangle $ after
performing the same operations above. To simplify our presentation, we will
not give a complete expression for $\left| \psi \left( \tau \right)
\right\rangle $ due to its complexity.

The fidelity is given by
\begin{eqnarray}
F &=&\left| \left\langle \psi _{id}\left( \tau \right) \right| \left. \psi
\left( \tau \right) \right\rangle \right| ^2  \nonumber \\
&=&1-2x\left( 1+p^2-q^2-r^2\right) +x^2\left[ \left( 1-q^2-r^2\right)
^2+2p^2\left( 1+q^2-r^2\right) +p^4\right] ,
\end{eqnarray}
where
\begin{eqnarray}
x=\left| \theta \right| ^2,\;p=\cos \varphi ,\;q=\frac s{2\sqrt{\Omega
_{12}^2+s^2/4}}\sin \varphi ,\;r=\frac{\Omega _{12}}{\sqrt{\Omega
_{12}^2+s^2/4}}\sin \varphi ,
\end{eqnarray}
with $s=g_b^2/\Delta _c$ and $\varphi =\pi \sqrt{\Omega _{12}^2+s^2/4}%
/\left( 2\Omega _{12}\right) $ .

Eq.~(11) shows that the fidelity $F$ is a function of $x$ $\in
\left[ 0,1\right] $. Thus, the average fidelity over all possible
two-qubit initial states is given by
\begin{eqnarray}
\overline{F} &=&\int_0^1F\left( x\right) dx  \nonumber \\
&=&\frac 13\left[ 1+p^4+q^4+r^2+r^4+p^2\left( -1+2q^2-2r^2\right) +q^2\left(
1+2r^2\right) \right] .
\end{eqnarray}

\begin{figure}[tbp]
\includegraphics[bb=41 233 464 512, width=8.6 cm, clip]{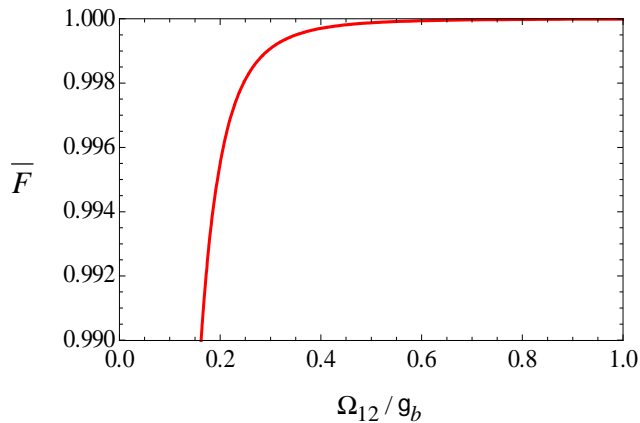} %
\vspace*{-0.08in}
\caption{Average fidelity $\overline{F}$ as a function of the Rabi frequency
$\Omega _{12}$ (in unit of $g_b$) for $\Delta _c=10g_b$. }
\label{fig:3}
\end{figure}

It can be verified that when the unwanted ``SQUID $b$''-resonator
off-resonant interaction in steps (ii) and (iv) is not considered (i.e., the
case for $g_b^2/\Delta _c=0$ or $s=0$), we have $p=q=0$ and $r=1,$ leading
to $F=1$ and $\overline{F}=1.$ We have plotted the average fidelity $%
\overline{F}$ for the case $\Delta _c=10g_b$ (Fig. 3). One can see from Fig.
3 that the average fidelity $\overline{F}$ increases as the Rabi frequency $%
\Omega _{12}$ of the pulse applied to SQUID $b$ becomes larger, and the $%
\overline{F}$ is $\sim 1$ when $\Omega _{12}=0.6g_b.$

\textit{Conclusion.}---We have presented a way to realize a two-qubit
controlled\ phase gate with two SQUIDs, by the use of a microwave
superconducting resonator. As shown above, in this proposal, (\textit{a})
SQUIDs, which often have considerable parameter nonuniformity, can be used; (%
\textit{b}) the adjustment of the level spacings, which is undesirable in
experiment, is avoided; and (\textit{c}) neither the adiabatic passage nor a
second-order detuning is needed and thus the gate can be performed much
faster.

\textit{Acknowledgments.}---S.H. was supported in part by the NSF Contract
No. DMR-0325551. S.L., J.Y.L and X.L.H. were supported in part by the
National Science Foundation of China under Grants 10904128 and 20804039.
C.P.Y acknowledges the funding support from the Hangzhou Normal University.

\end{document}